\def\year{2022}\relax
\documentclass[letterpaper]{article} 
\usepackage{aaai22}  
\usepackage{times}  
\usepackage{helvet}  
\usepackage{courier}  
\usepackage[hyphens]{url}  
\usepackage{graphicx} 
\usepackage{natbib}  
\usepackage{caption} 
\DeclareCaptionStyle{ruled}{labelfont=normalfont,labelsep=colon,strut=off} 
\usepackage{algorithm}
\usepackage{algorithmic}
\usepackage{ amssymb }
\usepackage{amsmath}
\usepackage{hyperref}
\usepackage[capitalize,noabbrev]{cleveref}
\usepackage{multirow}
\usepackage{makecell}
\usepackage{booktabs}
\usepackage{adjustbox}
\usepackage{caption}
\usepackage{newfloat}
\usepackage{listings}
\floatstyle{ruled}
\newfloat{listing}{tb}{lst}{}
\floatname{listing}{Listing}
\title{Avocodo: Generative Adversarial Network for Artifact-Free Vocoder}
\author{
    Taejun Bak\textsuperscript{\rm 1}\equalcontrib,
    Junmo Lee\textsuperscript{\rm 2}\footnotemark[1]\thanks{Work performed at NCSOFT.},
    Hanbin Bae\textsuperscript{\rm 3}\footnotemark[2],
    Jinhyeok Yang\textsuperscript{\rm 4}\footnotemark[2],
    Jae-Sung Bae\textsuperscript{\rm 3}\footnotemark[2],
    Young-Sun Joo\textsuperscript{\rm 1}
}
\affiliations{
    \textsuperscript{\rm 1}AI Center, NCSOFT, Seongnam, Korea\\
    \textsuperscript{\rm 2}SK Telecom, Seoul, Korea\\
    \textsuperscript{\rm 3}Samsung Research, Seoul, Korea\\
    \textsuperscript{\rm 4}Supertone Inc., Seoul, Korea

    happyjun@ncsoft.com, ljun4121@sk.com
}

\begin{document}

\maketitle

\begin{abstract}
Neural vocoders based on the generative adversarial neural network (GAN) have been widely used due to their fast inference speed and lightweight networks while generating high-quality speech waveforms. Since the perceptually important speech components are primarily concentrated in the low-frequency bands, most GAN-based vocoders perform multi-scale analysis that evaluates downsampled speech waveforms. This multi-scale analysis helps the generator improve speech intelligibility. However, in preliminary experiments, we discovered that the multi-scale analysis which focuses on the low-frequency bands causes unintended artifacts, e.g., aliasing and imaging artifacts, which degrade the synthesized speech waveform quality. Therefore, in this paper, we investigate the relationship between these artifacts and GAN-based vocoders and propose a GAN-based vocoder, called Avocodo, that allows the synthesis of high-fidelity speech with reduced artifacts. We introduce two kinds of discriminators to evaluate speech waveforms in various perspectives: a collaborative multi-band discriminator and a sub-band discriminator. We also utilize a pseudo quadrature mirror filter bank to obtain downsampled multi-band speech waveforms while avoiding aliasing. According to experimental results, Avocodo outperforms baseline GAN-based vocoders, both objectively and subjectively, while reproducing speech with fewer artifacts.
\end{abstract}

\section{Introduction}
Speech synthesis generates speech waveforms that correspond to the input text. An acoustic model initially generates acoustic features corresponding to the input text \citep{Tacotron, TransformerTTS, FastSpeech, FastSpeech2}. A vocoder then converts the acoustic features into a speech waveform \citep{World, Wavenet}. 
With the emergence of deep learning, neural vocoders can generate high-fidelity speech waveforms that are indistinguishable from human recordings \citep{WaveGlow, FloWaveNet, Parallel-Wavenet}.
Recently, vocoders based on generative adversarial network (GAN) \citep{GAN} with non-autoregressive convolutional architectures have been proposed \citep{Parallel-WaveGAN, MelGAN, VocGAN, HiFi-GAN, MB-MelGAN, StyleMelGAN, Fre-GAN}. Compared to other neural vocoders \citep{Wavenet, WaveGlow, FloWaveNet, Parallel-Wavenet}, GAN-based vocoders are faster and lighter while still maintaining a high level of synthesized speech quality. Specifically, a generator converts input features such as random noise or a mel-spectrogram into speech waveforms. A discriminator then evaluates the generated speech waveforms.

Because the speech spectrum in the low-frequency bands is much more crucial to perceptual quality, most GAN-based vocoders perform multi-scale analysis that evaluates the downsampled speech waveforms. Multi-scale analysis allows a generator to focus on the speech spectrum in low-frequency bands; downsampling limits the frequency range of speech by decreasing the sampling rate \citep{Nyq}. In MelGAN \citep{MelGAN}, a multi-scale discriminator (MSD) evaluates the downsampled waveforms that used an average pooling technique. In HiFi-GAN \citep{HiFi-GAN}, a multi-period discriminator (MPD) specializing in periodic components was proposed. It discriminates downsampled waveforms obtained by using an equally spaced sampling technique with various periods. Consequently, these GAN-based vocoders have successfully increased the quality of synthesized speech \citep{VocGAN, Fre-GAN, UnivNet}.

In preliminary experiments, however, we discovered that, GAN-based vocoders suffer from two major issues. The first issue is the artifacts caused by the upsampling layer \citep{imgartifact}. For example, artifacts in high-frequency bands degrade the quality of speech by introducing noise. The second issue is the degraded reproducibility of the harmonic components. The fundamental frequency (\textit{F}\textsubscript{0}) of synthesized speech is often inaccurate \citep{CARGAN} with the aliasing during simple downsampling, such as an average pooling or an equally spaced sampling, being one of the reasons behind this problem. These artifacts significantly reduce the perceptual quality, when synthesizing speech with large pitch variation \citep{ towards_robust, daft-exprt}. The preceding issues are analyzed further in the following section.

To address these issues, we propose Avocodo, a GAN-based vocoder that synthesizes high-quality speech waveform by minimizing artifacts. Avocodo is designed to factor in and suppress artifacts that should be considered in the  digital signal processing. The Avocodo contains two discriminators; a collaborative multi-band discriminator (CoMBD) and a sub-band discriminator (SBD). (1) The CoMBD comprises a novel structure for multi-scale analysis and suppressing upsampling artifacts. Since the CoMBD discriminates full-resolution waveform and intermediate outputs altogether, it takes two advantages. First, it helps the generator to focus on the spectral features in low-frequency bands by the multi-scale analysis. Second, the generator is trained to learn to suppress artifacts caused by the upsampling layer. 
(2) The SBD improves the sound quality by discriminating frequency-wise decomposed waveforms. It allows the generator to learn the speech spectrum, not only in low-frequency bands, but in high-frequency bands as well. In addition, to further improve the sound quality, we utilize a pseudo quadrature mirror filter bank (PQMF) \citep{PQMF} equipped with high stopband attenuation as a downsampling method, as opposed to the simple downsampling methods that cause aliasing, which are commonly used in conventional GAN-based vocoders.

We evaluated Avocodo's performance with both objective and subjective evaluations. The subjective evaluation shows that Avocodo can synthesize high-quality speech and be robust in unseen speaker synthesis. In addition, in the objective evaluation, accuracy in \textit{F}\textsubscript{0} reconstruction and the quality of high-frequency bands are improved. Finally, an analysis on the effect of alias-free methods on the artifacts is described.

\begin{figure}[t]
  \centerline{\includegraphics[width=1\linewidth]{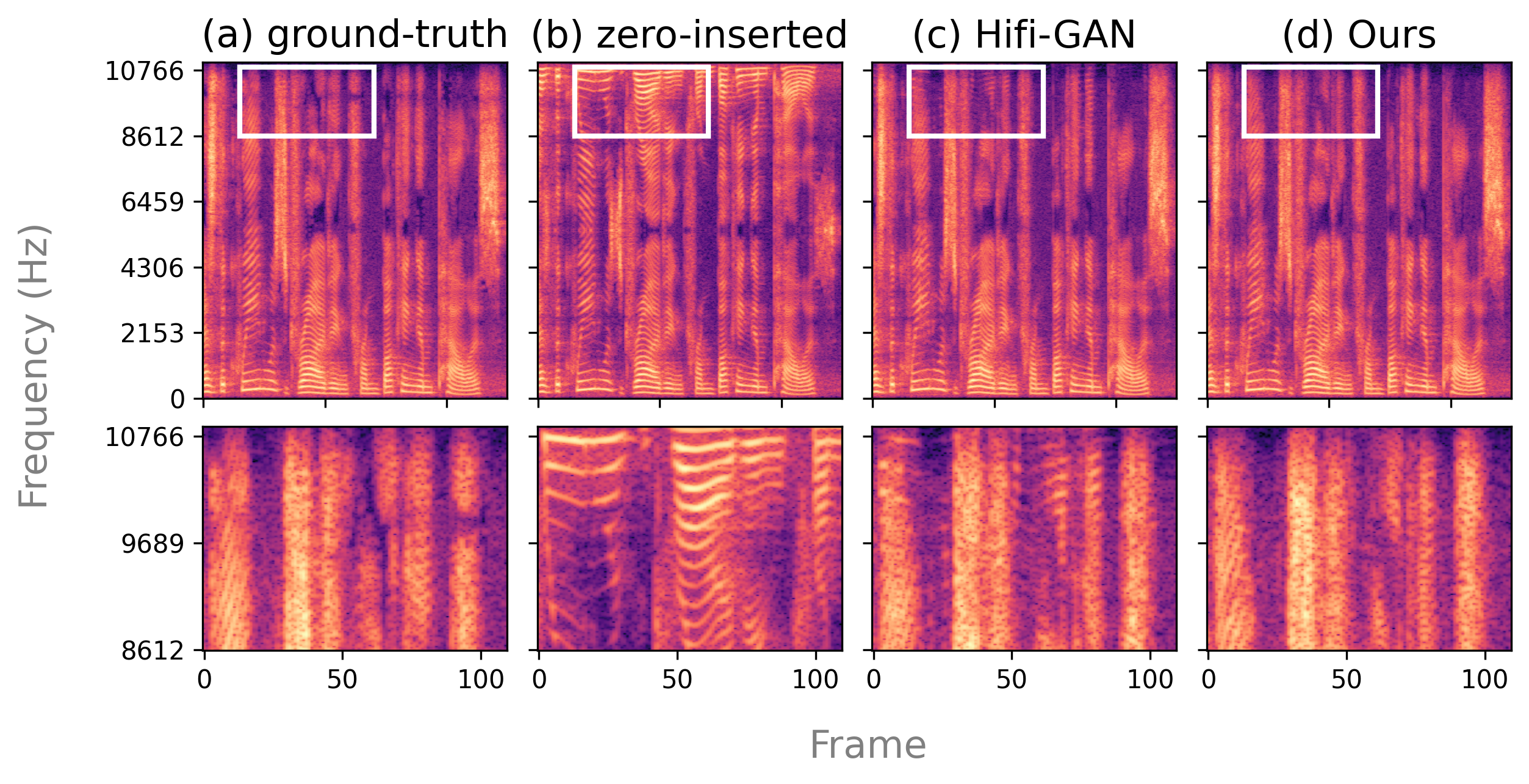}}
  \caption{Sub-figures in the first row show spectrograms of (a) a ground truth, generated waveforms from (b) a zero-stuffing, (c) HiFi-GAN, and (d) proposed method. Enlarged versions of the white rectangular boxes are depicted in the second row; note that mirrored low frequencies in (b) still exist in results from (c), but not from (d).}
  \label{fig:imaging_artifact}
\end{figure}
\section{Artifacts in GAN-based Vocoders}\label{Artifacts}
\subsection{Upsampling artifacts}\label{sec:ImageArtifact}
GAN-based vocoders incorporate upsampling layers to increase the rate of input features, such as a mel-spectrogram, up to the sampling rate of waveform \citep{MelGAN, HiFi-GAN, VocGAN}. However, upsampling layer, such as transposed convolution, causes several artifacts, tonal artifacts \citep{imgartifact} being one example. Tonal artifacts appears as a horizontal line on spectrogram. Additionally, mirrored low frequencies are observed in high-frequency bands, which are called \emph{imaging artifacts} in this paper.
In digital signal processing, the signal is upsampled by inserting zeros between neighboring samples, and then applying low-pass filtering \citep{interpolation}. Without the filtering, low-frequency components appear in high-frequency bands, as shown in \cref{fig:imaging_artifact}b, because the spectrum repeats over a cycle of sampling rate. The upsampling layer should also remove enough of the unintended frequency components, but it is unable to meet that criteria. As shown in \cref{fig:imaging_artifact}c, the unintended frequency components, which are imaging artifacts, eventually degrade the speech quality by distortion in high-frequency bands. Such artifacts are similar to texture sticking of image generative models, reported in \citep{StyleGAN_V3}.

To address these artifacts in GAN-based synthesis, several studies have proposed modifying the structure of the upsampling layer \citep{imgartifact, StyleGAN_V3, AdvAs, waveunet}. However, these methods either increase the model complexity or are insufficient in suppressing artifacts. Therefore in this study, a novel discriminator and loss functions that do not modify the upsampling layer are designed to suppress artifacts.

\begin{figure}[t]  \centerline{\includegraphics[width=0.85\linewidth]{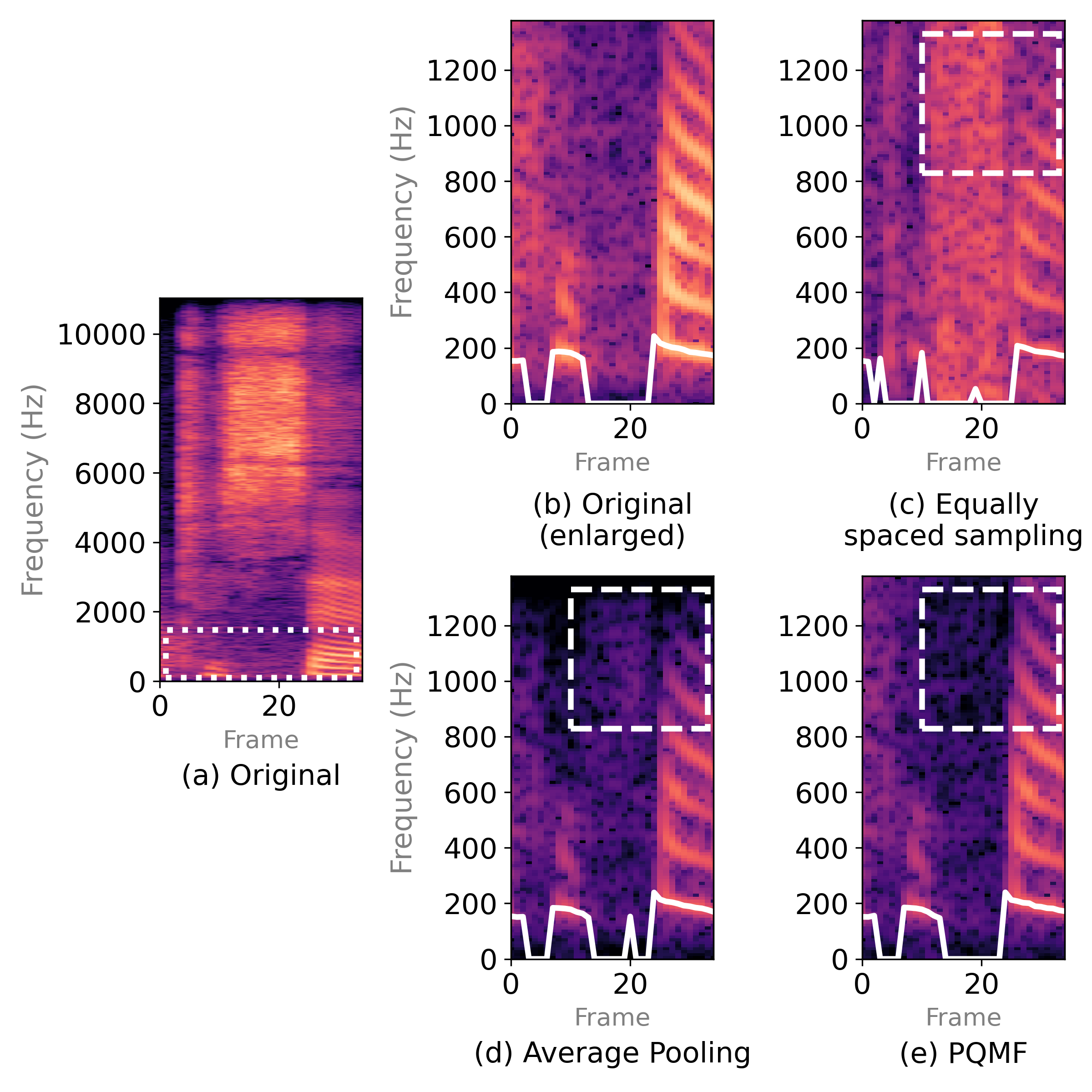}}
  \caption{Spectrograms of original and downsampled audio samples. Downsampling is performed for (a,b) the original waveform with (c) the equally spaced sampling, (d) the average pooling, and (e) PQMF. White solid lines are the \textit{F}\textsubscript{0} contours.}
  \label{fig:aliasing_ds}
\end{figure}
\begin{figure*}[t]
  \centerline{\includegraphics[width=0.9\linewidth]{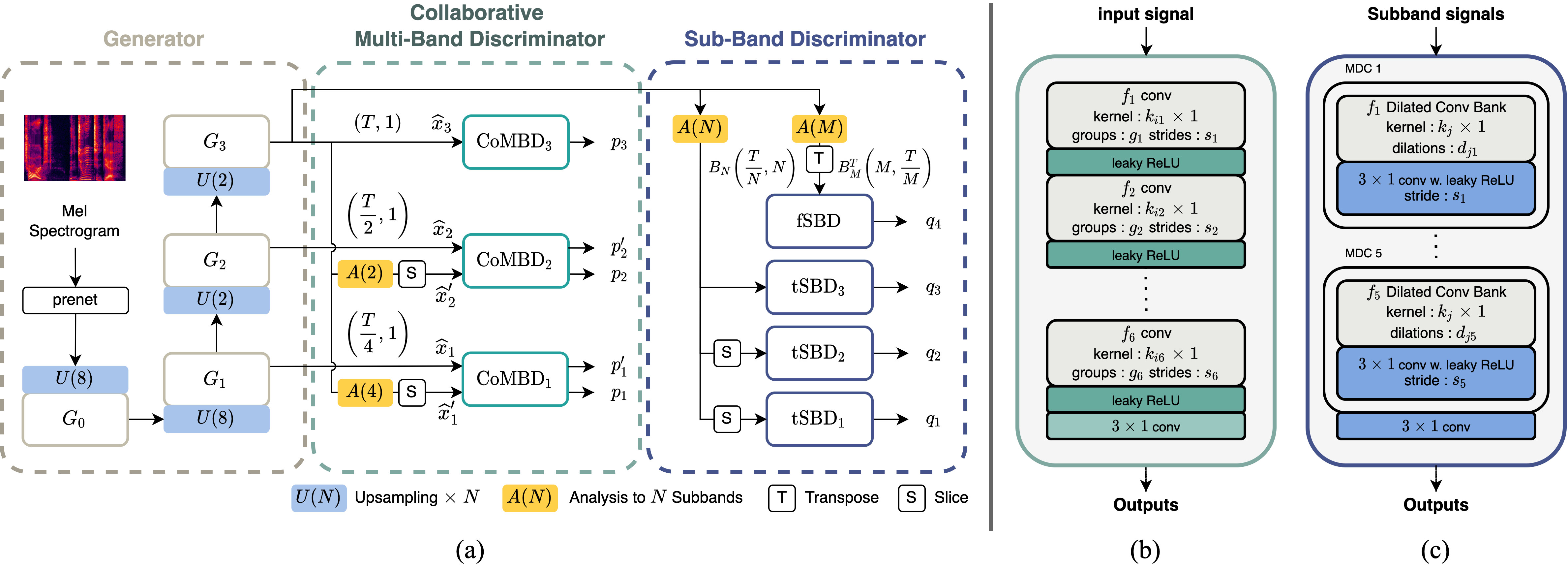}}
  \caption{Overall architecture of Avocodo (a). Avocodo comprises the generator and two discriminators: CoMBD and SBD. Detailed architectures of each sub-module of CoMBD and SBD are depicted in (b) and (c), respectively.}
  \label{fig:avocodo_system}
\end{figure*}
\subsection{Aliasing in downsampling}\label{Aliasing}
GAN-based vocoders use discriminators to evaluate downsampled waveforms to learn the spectral information in low-frequency bands. Typical downsampling methods, such as the average pooling used in \citep{MelGAN, HiFi-GAN, VocGAN} or the equally spaced sampling used in \citep{HiFi-GAN, Fre-GAN, UnivNet}, are easy to implement and efficient for obtaining band-limited speech waveforms. In preliminary experiments, however, aliasing was observed in the downsampled waveforms using the aforementioned methods.
\cref{fig:aliasing_ds} illustrates examples of the downsampled waveforms using several approaches; the downsampling factor is set to $8$. Taking into consideration downsampling uses equally spaced sampling (\cref{fig:aliasing_ds}c), high-frequency components, which are supposed to be removed, fold back and distort the harmonic frequency components at low-frequency bands. In the case of the average pooling (\cref{fig:aliasing_ds}d), which is a composition of a simple low-pass filtering and a decimation, aliasing is not as apparent in low-frequency bands but harmonic components over $800$Hz are distorted. As the downsampling factor increases, the artifacts increase too. Using these distorted downsampled waveforms during the training makes it difficult for the model to generate accurate waveforms.

To avoid this aliasing, downsampling using a band-pass filter equipped with a high stopband attenuation is required. PQMF, a digital filter, satisfies this requirement \citep{DurIAN, MB-MelGAN}. As shown in \cref{fig:aliasing_ds}d, downsampling using the PQMF preserves the harmonics well.

\section{Proposed Method}
\cref{fig:avocodo_system}a describes the overall architecture of Avocodo. It consists of a single generator and the two proposed discriminators. Taking a mel-spectrogram as input, the generator outputs not only full-resolution waveforms but also intermediate outputs. Subsequently, the CoMBD discriminates the full-resolution waveform and its downsampled waveforms along with the intermediate outputs; the PQMF is used as a low-pass filter to downsample the full-resolution waveform. Additionally, the SBD discriminates sub-band signals obtained by the PQMF analysis.

\subsection{Generator} 
The generator has the same structure as the HiFi-GAN generator, but it produces multi-scale outputs that is composed of both high-resolution and intermediate waveforms. The generator has four sub-blocks, three of which $G_k\space (1 \le k \le 3)$ generate waveforms $\hat{x}_k$ with the corresponding resolution of $\frac{1}{2^{3-k}}$ of the full resolution. To elaborate, $\hat{x}_3$ is a full-resolution waveform; moreover, $\hat{x}_1$ and $\hat{x}_2$ denote intermediate outputs. Each sub-block comprises multi-receptive field fusion (MRF) blocks \citep{HiFi-GAN} and transposed convolution layers. The MRF blocks contain multiple residual blocks of diverse kernel sizes and dilation rates to capture the spatial features of input. Additional projection layers are added, unlike HiFi-GAN, after each sub-block to return the intermediate outputs. Please note that our approach can be applied to any GAN-based vocoder using upsampling layers. In this paper, HiFi-GAN's generator is selected due to its acceptable performance.

\begin{figure*}[t]
  \centerline{\includegraphics[width=0.75\linewidth]{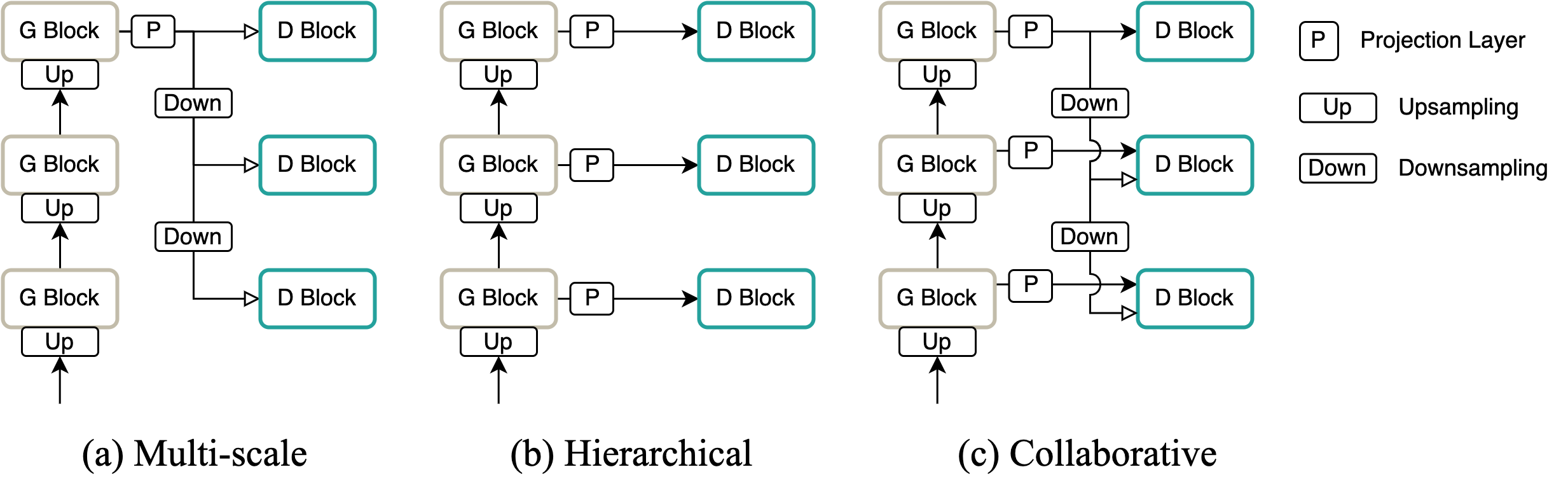}}
  \caption{Comparison of various structures for discriminators.}
  \label{fig:mbd-comparison}
\end{figure*}
\subsection{Collaborative Multi-Band Discriminator}
The proposed CoMBD discriminates multi-scale outputs from the generator. It comprises identical sub-modules, which evaluate waveforms at different resolutions. Additionally, each sub-module is based on the discriminator module of MSD. The module comprises fully convolutional layers and a leaky ReLU activation function.

Either a multi-scale structure (\cref{fig:mbd-comparison}a) or a hierarchical structure (\cref{fig:mbd-comparison}b) is commonly used in conventional GAN-based neural vocoders; however, in this paper, the two structures are combined to take advantage of each structure, as shown in \cref{fig:mbd-comparison}c. This collaborative structure helps the generator to synthesize high-quality waveforms with reduced artifacts.

The multi-scale structure increases speech quality by discriminating not only the full-resolution waveform but also the downsampled waveform \citep{MelGAN, HiFi-GAN, MB-MelGAN, UnivNet}. In particular, the discrimination of waveforms downsampled into multiple scales helps the generator to focus on the spectral features in low-frequency bands \citep{MelGAN}. Meanwhile, the hierarchical structure uses intermediate output waveforms of each generator sub-block, helping the generator to learn the various levels of acoustic properties in a balanced manner \citep{VocGAN, photographic}. In particular, the generator sub-blocks are trained to learn expansion and filtering in a balanced way by inducing the sub-blocks of the generator to generate a band-limited waveform. Therefore, upsampling artifacts are expected to be suppressed by adopting the hierarchical structure.

For the proposed collaborative structure, the sub-modules at low resolution, i.e., CoMBD$_1$ and CoMBD$_2$, take both the intermediate outputs $\hat{x}$ and the downsampled waveforms $\hat{x}^\prime$ as their inputs. For each resolution, both inputs share the sub-module. For example, as shown in \cref{fig:avocodo_system}, the intermediate output $\hat{x}_2$ and the downsampled waveform $\hat{x}_2^\prime$ share the weights of CoMBD$_2$ for output $p_2$ and $p_2^\prime$, respectively. The intermediate output and downsampled waveform are intended to match each other after collaboration. Note that no additional parameters are necessary for collaborating the two structures because of the weight-sharing process \citep{CoGAN}.

To further improve speech quality by reducing artifacts, a differentiable PQMF is adopted to obtain downsampled waveform with restricted aliasing. First, a full-resolution speech waveform is decomposed into $K$ sub-band signals $B_K$ by using the PQMF analysis. The $B_K$ comprise single-band signals $b_1,...,b_K$ with a length of $\frac{T}{K}$, where $T$ is the length of the full-resolution waveform \citep{PQMF}. Next, the first sub-band signal $b_1$ is selected corresponding to the lowest frequency band.

\subsection{Sub-Band Discriminator}
An SBD is introduced that discriminates multiple sub-band signals by PQMF analysis. The PQMF enables the $n^{th}$ sub-band signal $b_n$ to contain frequency information corresponding to the range from $(n-1)f_{s}/2N$ to $nf_{s}/2N$, where $f_{s}$ is the sampling frequency and $N$ is the number of sub-bands. Inspired by this characteristic of sub-band signals, the SBD sub-modules learn various discriminative features by using different ranges of the sub-band signals.

Two types of sub-modules are designed: one captures the changes in spectral feature over the time axis and the other captures the relationship between each sub-band signal. These two sub-modules are referred to as $\text{tSBD}$ and $\text{fSBD}$, respectively, in \cref{fig:avocodo_system}a. The $\text{tSBD}$ takes $B_N$ as its input and performs time-domain convolution with $B_N$. By diversifying sub-band ranges, each sub-module can be designed to learn the characteristics of a specific frequency range. In other words, $\text{tSBD}_{k}$ takes a certain number of sub-band signals $b_{i_k:j_k}$ as its input. In contrast, $\text{fSBD}$ takes the transposed version of $M$ channel sub-bands ${B_M}^T$. The composition of $\text{fSBD}$ is inspired by the spectral features of speech waveform, such as harmonics and formants.

Each sub-module of the SBD comprises stacked multi-scale dilated convolution banks \citep{NeuralPhoto} to evaluate sub-band signals. The dilated convolution bank contains convolution layers with different dilation rates that cover diverse receptive fields. Moreover, the SBD architecture follows an inductive bias as an accurate analysis on speech waveforms requires various receptive fields for each frequency range. Consequently, different dilation factors are prepared for each sub-module. 

Several neural vocoders utilize filter-banks to decompose speech waveforms and utilize discriminators to inspect the sub-band signals \citep{MB-MelGAN, StyleMelGAN, Fre-GAN}. In particular, the SBD is similar to the filter-bank random window discriminators (FB-RWDs) of the StyleMelGAN \citep{StyleMelGAN} as both obtain sub-band signals using the PQMF. However, SBD and FB-RWDs are considerably different. Each sub-module of SBD evaluates a different range of sub-band signals, whereas the FB-RWDs vary the number of sub-band signals for each discriminator. In addition, the SBD has many types of blocks: blocks used to observe a lower frequency band, a whole range of frequency bands, and a relationship between frequency bands. Consequently, the SBD can evaluate signals more effectively than FB-RWDs.

\subsection{Training Objectives}
\subsubsection{GAN Loss} For training GAN networks, the least square adversarial objective \citep{LSGAN} is used, which replaces a sigmoid cross-entropy term of the GAN training objective proposed in \citep{GAN} with the least square for stable GAN training. The GAN losses, $V$ for multi-scale outputs and $W$ for downsampled waveforms, are defined as follows:
\begin{equation}
    V(D_k;G)=\mathbb{E}_{(x_k,s)}\Big[(D_k(x_k)-1)^2+(D_k(\hat{x}_k))^2\Big] \label{eq:1} 
\end{equation}

\begin{align}
V(G;D_k)&=\mathbb{E}_{s}\Big[(D_k(\hat{x}_k)-1)^2\Big] \label{eq:2}\\
W(D_k;G)&=\mathbb{E}_{(x_k,s)}\Big[(D_k(x_k)-1)^2+(D_k(\hat{x}^\prime_k))^2\Big] \label{eq:3}\\
W(G;D_k)&=\mathbb{E}_{s}\Big[(D_k(\hat{x}^\prime_k)-1)^2\Big], \label{eq:4}
\end{align}

where $x_k$ represents the $k^{th}$ downsampled ground-truth waveform, and $s$ denotes the speech representation. In this paper, mel-spectrogram is utilized.

\subsubsection{Feature Matching Loss} Feature matching loss is a perceptual loss for GAN training \citep{ImpGAN}, which has been used in GAN-based vocoder systems \citep{MelGAN, HiFi-GAN,VocGAN}. Moreover, the feature matching loss of a sub-module in the discriminator can be established with L1 differences between the intermediate feature maps of the ground-truth and predicted waveforms. The loss can be defined as follows:
\begin{equation}\label{fm_loss}
    L_{fm}(G;D_t)=\mathbb{E}_{(x,s)}\Big[\sum^{T}_{t=1}{\frac{1}{N_t}}||D_t(x)-D_t(\hat{x})||\Big],
\end{equation}
where $T$ denotes the number of layers in a sub-module. $D_t$ and $N_t$ represent the $t^{th}$ feature map and the number of elements in feature map, respectively.

\subsubsection{Reconstruction Loss} Reconstruction loss based on a mel-spectrogram increases the stability and efficiency in the training of waveform generation \citep{Parallel-WaveGAN}. For that, L1 differences are calculated between the mel-spectrograms of the ground-truth $x$ and predicted $\hat{x}$ speech waveforms. The reconstruction loss can be expressed as follows:
\begin{equation}\label{spec_loss}
    L_{spec}(G)=\mathbb{E}_{(x,s)}\Big[||\phi(x)-\phi(\hat{x})||_{1}\Big],
\end{equation}
where $\phi(\cdot)$ denotes the transform function to mel-spectrogram.

\subsubsection{Final Loss} Final loss for the overall system training can be established from the aforementioned loss terms and defined as follows:
\begin{align}
    \begin{split}
    L^{total}_D&=\sum^{P}_{p=1} V(D^C_p;G)\\
    &\;\;\;+\sum^{P-1}_{p=1} W(D^C_p;G)+\sum^{Q}_{q=1} V(D^S_q;G) \label{floss_eq:1}\\ 
    \end{split}\\
    \begin{split}
    L^{total}_G&=\sum^{P}_{p=1} \Bigg[V(G;D^C_p)+\lambda_{fm}L_{fm}(G;D^C_p)\Bigg] \\
    &\;\;\;\;+\sum^{P-1}_{p=1} \Bigg[W(G;D^C_p) + \lambda_{fm}L_{fm}(G;D^C_p)\Bigg]\\
    &\;\;\;\;+\sum^{Q}_{q=1} \Bigg[V(G;D^S_q) + \lambda_{fm}L_{fm}(G;D^S_q)\Bigg]\\
    &\;\;\;\;+\lambda_{spec}L_{spec}(G), \label{floss_eq:2}
    \end{split}
\end{align}

where $D^C_p$ and $D^S_q$ denote the $p^{th}$ sub-module of CoMBD and the $q^{th}$ sub-module of SBD, respectively. $\lambda_{fm}$ and $\lambda_{spec}$ denote the loss scales for feature matching and reconstruction losses, respectively. $\lambda_{fm}$ and $\lambda_{spec}$ are set as $2$ and $45$, respectively.

\begin{table*}[t]
    \centering
    \begin{tabular}{cccccccc}
        \toprule
        \multirow{2}{*}{Model} & \multicolumn{3}{c}{MOS (CI)} & \multirow{2}{*}{\thead{\# G Param \\ (M)}} &
        \multirow{2}{*}{\thead{\# D Param \\ (M)}} & 
        \multirow{2}{*}{\thead{Inference \\ Speed (CPU)}} &
        \multirow{2}{*}{\thead{Inference \\ Speed (GPU)}} \\ \cmidrule{2-4}
         & LJ & Unseen(EN) & Unseen(KR) \\
        \midrule
        Ground Truth     & 4.362($\pm$0.07) & 4.173($\pm$0.08) & 4.690($\pm$0.05) & {-} & {-} & {-} & {-} \\
        VocGAN      & 4.135($\pm$0.08) & 3.638($\pm$0.09)  & 3.770($\pm$0.07) & {7.06} & {12.03} & {18.20x} & {235.00x} \\
        StyleMelGAN & 3.663($\pm$0.09) & 3.597($\pm$0.09) & 1.990($\pm$0.08) & {3.55} & {5.90} & {14.33x} & {180.09x} \\
        HiFi-GAN $V$1 & 4.150($\pm$0.08) & 3.940($\pm$0.09) & 3.810($\pm$0.06)  & {13.94} & {70.72}  & {15.95x} & {157.54x} \\
        Avocodo $V$1  & \textbf{4.258($\pm$0.08)} & \textbf{4.080($\pm$0.08)} &  \textbf{3.972($\pm$0.07)} & {13.94} & {27.07} & {15.45x} & {156.21x} \\
        \bottomrule
    \end{tabular}
    \caption{MOS results with 95\% CI, the number of parameters and inference speed of CPU and GPU.}
    \label{tab:MOS}
\end{table*}

\begin{table*}[t]
\begin{adjustbox}{width=1\textwidth}
\small
    \centering
    \begin{tabular}{crrrrrrrr}
        \toprule
        \multicolumn{9}{c}{Single speaker speech synthesis}  \\
        \midrule
        Model & \textit{F}\textsubscript{0} RMSE($\downarrow$) & \textit{F}\textsubscript{0} AE-STD($\downarrow$) &  VUV\textsubscript{fpr}($\downarrow$) & VUV\textsubscript{fnr}($\downarrow$) & MCD($\downarrow$) & PESQ($\uparrow$) & LSD-LF($\downarrow$) & LSD-HF($\downarrow$) \\
        \hline
        VocGAN & 37.51 & 38.19 & 20.15 & 12.45 & 2.63 & 3.25 & 7.61 & 9.50 \\
        StyleMelGAN & 36.60 & 36.42 & 19.78 & 14.12 & 3.82 & 2.30 & 8.61 & 10.14 \\
        HiFi-GAN $V$1  & 35.96 & 37.11 & 18.67 & 11.13 & 2.25 & 3.64 & 7.05 & 9.72 \\
        Avocodo $V$1  & \textbf{33.98} & \textbf{34.97}  & \textbf{17.74} &  \textbf{10.12} & \textbf{2.06} & \textbf{3.81} & \textbf{6.90} & \textbf{9.13} \\
        \bottomrule
    \end{tabular}
    \end{adjustbox}
    \caption{Objective evaluations results for the single speaker speech synthesis task.}
    \label{tab:objective_evaluation}
\end{table*}

\section{Experimental Setup}\label{experiments}
\subsection{Datasets}
\subsubsection{Single speaker speech synthesis} The LJSpeech \citep{LJ} dataset was used for a single speaker experiment. The dataset contains $13{,}100$ audio samples recorded by a native English-speaking female speaker, which amounted to a total recording time of $24$h and the audio samples are sampled at $22{,}050$Hz with $16$bit. For the testset, $150$ samples are randomly selected.

\subsubsection{Unseen speaker speech synthesis}
Public English dataset, i.e., VCTK \citep{vctk}, (Unseen(EN)) and internal Korean dataset (Unseen(KR)) were used to evaluate the generalization of the proposed model. VCTK consists of audio samples recorded by $109$ speakers, and the total amount of samples is $44$h long. $9$ speakers were selected for the testset. The internal Korean dataset contains $156$ speakers, amounting to an approximately $244$h long recording. Among them, $16$ unseen speakers were excluded from the training. The voice style of dataset comprises a variety of reading, daily conversations, and acting. The audio samples of VCTK and internal datasets were resampled at $24{,}000$Hz and $22{,}050$Hz, respectively.

\subsection{Training Setup}
As the baseline models, we selected HiFi-GAN\footnote{\url{https://github.com/jik876/hifi-gan}}, VocGAN, and StyleMelGAN\footnote{\url{https://github.com/kan-bayashi/ParallelWaveGAN}}. HiFi-GAN utilizes discriminators based on multi-scale structure downsampling with average pooling and equally spaced sampling. VocGAN uses a discriminator based on hierarchical structure downsampling with average pooling. StyleMelGAN utilizes discriminators that discriminate the sub-band signals of random window selected signal obtained by PQMF analysis.
For the single speaker speech synthesis, Avocodo\footnote{Source code is available at \url{https://github.com/ncsoft/avocodo}.} and HiFi-GAN were both trained up to $3$M steps. VocGAN and StyleMelGAN were trained up to $2.5$M and $1.5$M steps, respectively. Next, for the unseen speaker synthesis, all models were trained up to $1$M steps.

The hyper-parameters of Avocodo's generator are the same as that of the HiFi-GAN. The HiFi-GAN generator has two versions with an identical architecture but different number of parameters: $V$1 is larger than $V$2, and Avocodo also follows this rule. The number of sub-bands $N$ is $16$ for tSBD and $M$ is $64$ for fSBD. Moreover, the parameters of PQMF were selected empirically. An AdamW optimizer \citep{AdamW} was used with an initial learning rate of $0.002$. The optimizer parameters $(\beta_1, \beta_2)$ were set as $(0.8, 0.99)$, and an exponential learning rate decay of $0.999$ was applied \citep{HiFi-GAN}.

Input features are ground-truth mel-sepctrogram extracted from recorded speech waveform. $80$ bands of mel-spectrograms were calculated from audio samples using the short-time Fourier transform (STFT). The STFT parameters for $22{,}050$Hz were set as $1{,}024$, $1{,}024$, $256$ for the number of STFT bin, window sizes, and hop sizes, respectively. For $24$kHz, the parameters were set as $2{,}048$, $1{,}200$, $300$, respectively. Each audio sample was sliced with the random window selection method. The segment size was $8{,}192$, which is about $0.4$s long.

\section{Experimental Results}
\subsection{Audio Quality \& Comparison} \label{AudioQuality}
The performance of the proposed model for each dataset was assessed using various subjective and objective measurements\footnote{Audio samples are available at \url{https://nc-ai.github.io/speech/publications/Avocodo}.}.

\subsubsection{Subjective evaluation} 5-scale mean opinion score (MOS) tests were conducted for single and unseen speaker syntheses. For the English dataset, $15$ native English speakers and $19$ native Korean speakers participated for the Korean dataset. All participants were requested to assess the sound quality of $20$ audio samples randomly selected from each testset.

\cref{tab:MOS} lists subjective evaluation results. We can see Avocodo $V$1 performs the best performance in both single and unseen speaker synthesis tasks. For synthesizing high-quality speech waveform of unseen speakers, learning generalized characteristic of speech signals is crucial. Because artifacts inhibit the generator from learning the generalized characteristics of speech signals, Avocodo's approaches for suppressing artifacts are much more robust than baseline models. In particular, Avocodo outperforms baseline models even in Unseen(KR). Since the dataset includes various speech styles, the overall differences from the ground truth are larger than other datasets. Note that StyleMelGAN even failed to train for Unseen(KR).

\subsubsection{Objective evaluation} Objective evaluations were conducted to quantitatively compare vocoders. To validate the reproducibility of \textit{F}\textsubscript{0}, we measured the \textit{F}\textsubscript{0} root mean square error (\textit{F}\textsubscript{0} RMSE) in the voiced frame. Because the artifacts exist in very short regions (only a few frames), the average value of \textit{F}\textsubscript{0} RMSE is insufficient to represent the artifacts. Therefore, we further calculated the standard deviation of \textit{F}\textsubscript{0} absolute error (\textit{F}\textsubscript{0} AE-STD); a low \textit{F}\textsubscript{0} AE-STD value means less distortion exists in harmonics. For evaluating the accuracy in voiced/unvoiced (VUV) frame, false positive and negative rates of the VUV classification (VUV\textsubscript{fpr}, VUV\textsubscript{fnr}) were measured. To measure the perceived quality of the synthesized speech, the mel-cepstral distortion (MCD)\citep{MCD} and perceptual evaluation of speech quality (PESQ)\citep{PESQ} were calculated. Additionally, we measured the log-spectral distance \citep{lsd, nuwave2} in low-frequency bands from $0$Hz to $5.5$kHz (LSD-LF) and in high-frequency bands from $5.5$kHz to $11.02$kHz Hz (LSD-HF); the low value of LSD-HF means the imaging artifacts are less.

\cref{tab:objective_evaluation} shows that Avocodo $V$1 also outperformed baseline models in overall results. In particular, due to the methods for suppressing upsampling artifacts, LSD-HF results show that Avocodo improves reproducibility in  high-frequency bands. Avocodo also takes advantage of reduced aliasing in training. Training with aliased waveform makes it easy to distort harmonic components. Because of anti-aliasing methods of Avocodo, \textit{F}\textsubscript{0} AE-STD and VUV errors are improved.
Despite the smaller number of parameters in discriminators, Avocodo also performs better than HiFi-GAN. Inference times for these two models are almost the same in CPU (Intel i7 CPU 3.00GHz) and single-GPU (NVIDIA V100) environments.

\subsection{Discriminator-wise Comparison}
MOS test was conducted for single speaker synthesis task to compare the performances of the proposed discriminators, MSD of MelGAN and MPD of HiFi-GAN. All discriminators were trained with Avocodo's generator $V2$. $\lambda_{fm}$ and $\lambda_{spec}$ are observed to empirically affect the training, therefore they were adjusted to $2$ and $10$, respectively. However, $\lambda_{spec}$ was adjusted to $20$ for the CoMBD which has a larger loss value owed to weight-sharing. \cref{tab:mos_ablation} shows that each Avocodo discriminator contributes to the generator synthesizing higher-quality speech with fewer artifacts. 

\begin{figure}[t]
  \centerline{\includegraphics[width=0.9\linewidth]{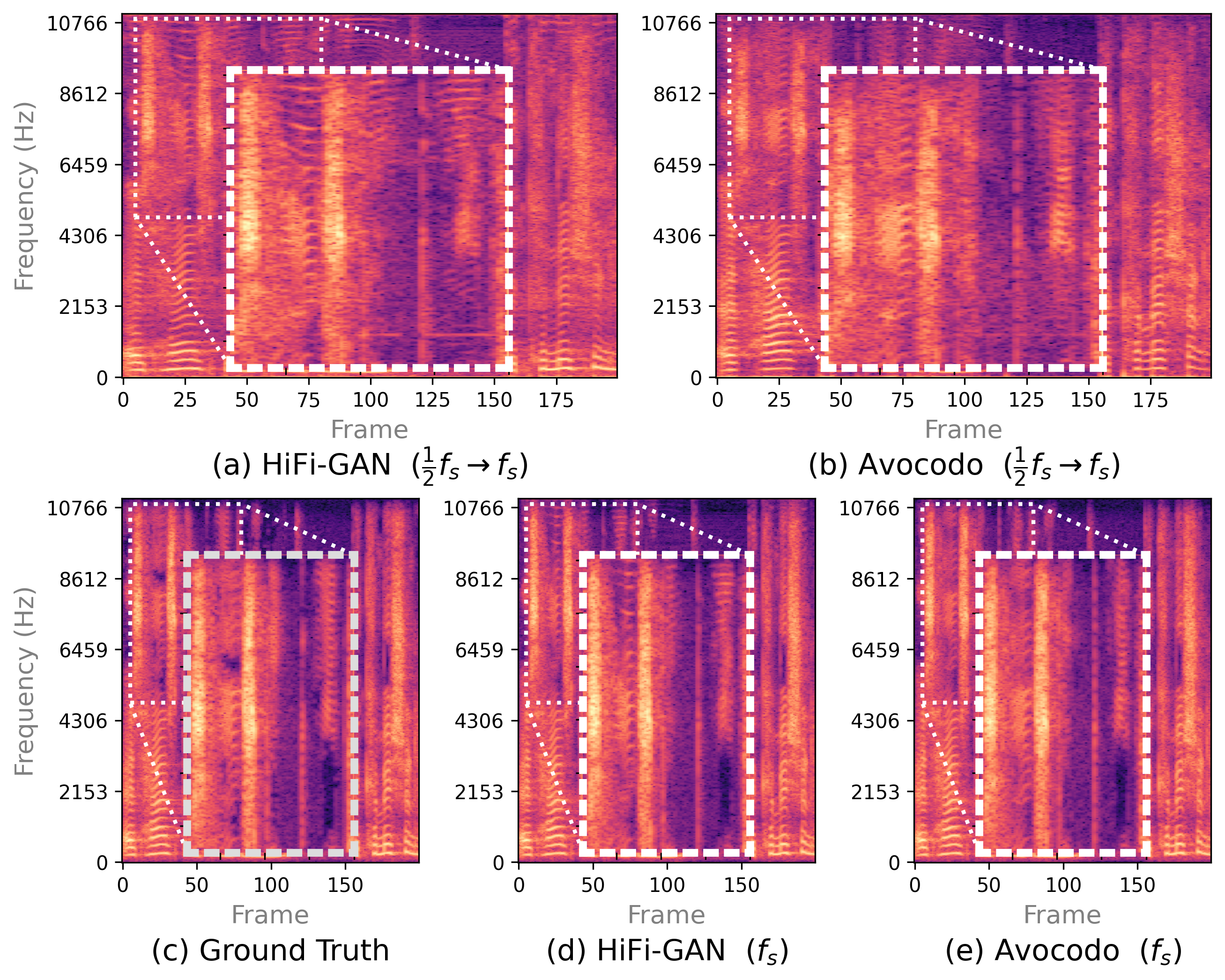}}
  \caption{Linear-scale spectrograms of synthesized audio samples. In case of HiFi-GAN, artifacts from upsampling layer in (a) remain and distort final outputs as shown in (d). Meanwhile,  artifacts are suppressed in (b) and final outputs (e).}
  \label{fig:upsampling_artifact}
\end{figure}
\begin{table}[t]
    \centering
    \begin{tabular}{cc}
        \toprule
        Model & MOS (CI) \\
        \midrule
        MSD\citep{MelGAN} & 3.743 ($\pm$0.07)  \\
        MPD\citep{HiFi-GAN} & 3.675 ($\pm$0.08)  \\
        CoMBD     & 4.156 ($\pm$0.05)  \\
        SBD     & 4.130 ($\pm$0.06)   \\
        \bottomrule
    \end{tabular}
    \caption{MOS results of the discriminator-wise comparison with 95\% CI.}
    \label{tab:mos_ablation}
\end{table}
\subsection{Analysis on artifacts}
\subsubsection{Upsampling Artifacts} The ability of Avocodo to suppress artifacts is explained by observing the upsampling artifacts occurring in intermediate upsampling layers of the generator. Audio samples are generated with HiFi-GAN and Avocodo, and their linear-scale spectrograms are depicted in \cref{fig:upsampling_artifact}. Audio samples of the first row of \cref{fig:upsampling_artifact} are the projected output of the last transposed convolution layer for upsampling from $\frac{1}{2}f_{s}$ to $f_{s}$ while skipping the last MRF block; \cref{fig:upsampling_artifact}a and \cref{fig:upsampling_artifact}b correspond to samples from HiFi-GAN and Avocodo, respectively.
Meanwhile, samples of the second row of \cref{fig:upsampling_artifact} are obtained by the complete generator. Tonal and imaging artifacts caused by transpose convolution exist in audio samples from HiFi-GAN as shown in \cref{fig:upsampling_artifact}a. Imaging artifacts still remain in the final output as shown in \cref{fig:upsampling_artifact}d. However, Avocodo's generator learns to remove artifacts occurring in intermediate upsampling layers from CoMBD. Therefore, no artifacts are present in neither \cref{fig:upsampling_artifact}b nor \cref{fig:upsampling_artifact}e.

\subsubsection{Aliasing} 
\begin{figure}[t]
  \centerline{\includegraphics[width=0.9\linewidth]{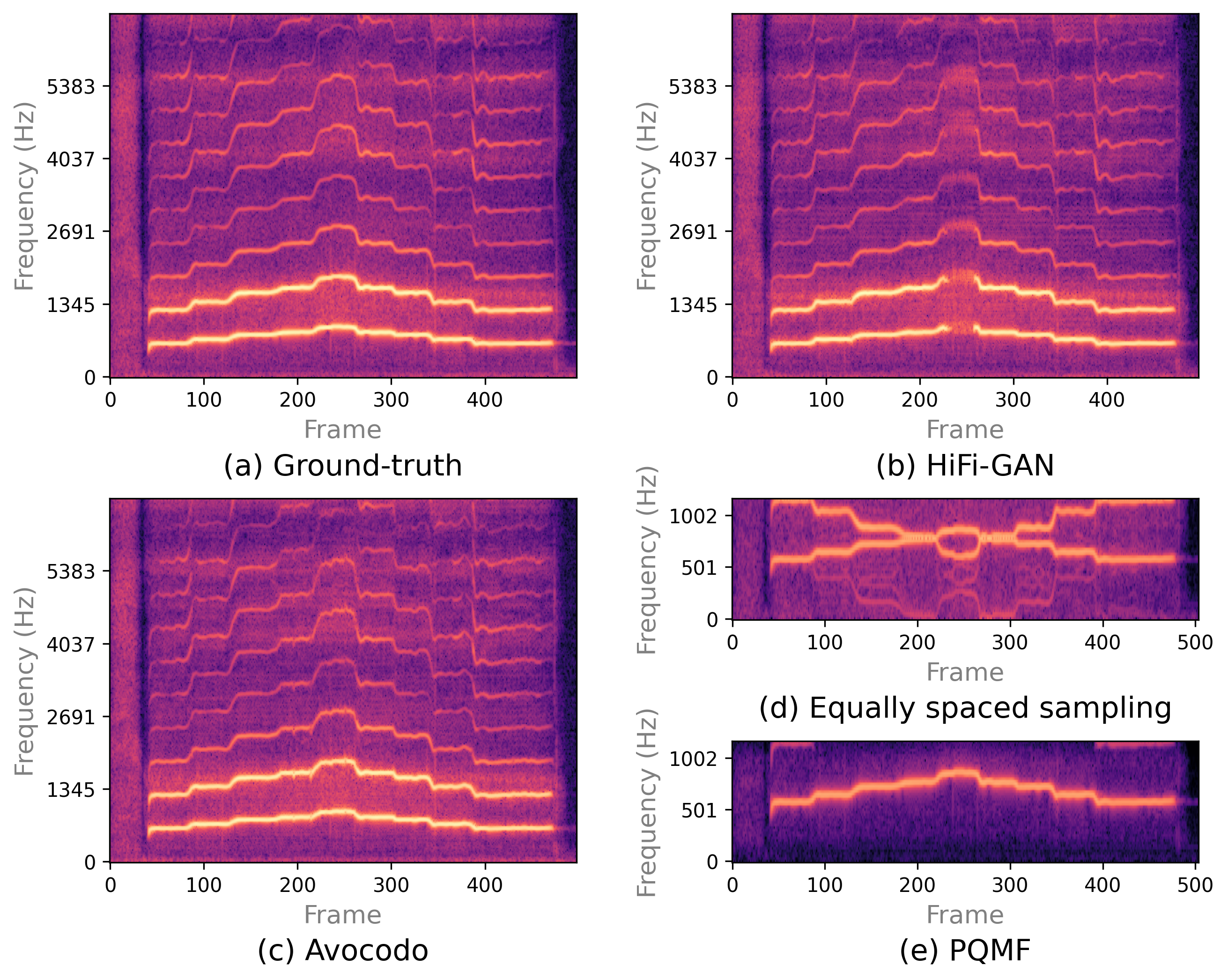}}
  \caption{Examples of failed \textit{F}\textsubscript{0} reconstruction. Due to aliasing of downsampled waveform in (d), HiFi-GAN fails to synthesize high \textit{F}\textsubscript{0} over $750$Hz (b).}
  \label{fig:aliasing_singing}
\end{figure}

To observe the distortion in \textit{F}\textsubscript{0} caused by aliasing, we trained GAN-based vocoders with singing voice datasets with a large range of \textit{F}\textsubscript{0}.
Large-scale downsampling for adequately modeling low-frequency components causes incomplete \textit{F}\textsubscript{0} reconstruction; for example, HiFi-GAN and VocGAN downsample by a factor of up to 11 and 16, respectively. In \cref{fig:aliasing_singing}d, the harmonic components of the downsampled waveforms are distorted due to the aliasing caused by downsampling, while downsampled waveforms with anti-aliasing PQMF preserve \textit{F}\textsubscript{0} as shown in \cref{fig:aliasing_singing}e. Therefore, HiFi-GAN (\cref{fig:aliasing_singing}b), trained using distorted downsampled waveforms, fails to reconstruct \textit{F}\textsubscript{0} higher than $750$Hz, while Avocodo (\cref{fig:aliasing_singing}c) preserves \textit{F}\textsubscript{0} contour.

\section{Conclusions} \label{conclusions}
In this paper, an artifact-free GAN-based vocoder, Avocodo, is proposed. Artifacts, such as upsampling artifacts and aliasing, are observed to originate from the limitation of the upsampling layer and the objective function biased towards the low-frequency bands obtained by naive downsampling methods. To solve these problems, two novel discriminators, namely CoMBD and SBD, are designed. The CoMBD performs multi-scale analysis with a collaborative structure of multi-scale and hierarchical structures. The SBD discriminates the sub-band signals decomposed by PQMF analysis in both time and frequency aspects. Furthermore, PQMF is utilized for downsampling and PQMF analysis. Various experimental results proved that these discriminators and the PQMF effectively reduce the artifacts in synthesized speech.

\bibliography{ref}

\newpage
\onecolumn
\appendix
\section{Detailed architecture}

\subsection{Collaborative Multi-Band Discriminator} \label{MBD}
The architecture of the sub-module of the CoMBD is based on a discriminator block of the multi-scale discriminator (MSD) in MelGAN \citep{MelGAN}. However, we set the hyper-parameters of the sub-modules of the CoMBD different from the MelGAN. The sub-modules are equipped with different receptive fields for each resolution unlike MSD. Detailed hyper-parameters of CoMBD are described in \cref{tab:appendix:CoMBD-params}. Each sub-discriminator module is a Markovian window-based discriminator that can be established using strided convolutional layers \citep{MelGAN}. Additionally, grouped convolution layers are also used to maintain the number of parameters while increasing the number of filters of the convolutional layer. 
\begin{table}[t]
\centering
      \begin{tabular}{lccc}
        \toprule
             & CoMBD\textsubscript{1} & CoMBD\textsubscript{2} & CoMBD\textsubscript{3} \\
        \midrule
        Kernel size  & [7, 11, 11, 11, 11, 5] & [11, 21, 21, 21, 21, 5] &  [15, 41, 41, 41, 41, 5]   \\
        Kernel filters & \multicolumn{3}{c}{[16, 64, 256, 1024, 1024, 1024]}      \\
        Kernel groups   & \multicolumn{3}{c}{[1, 4, 16, 64, 256, 1]} \\
        Kernel stride  & \multicolumn{3}{c}{[1, 1, 4, 4, 4, 1]}  \\
        \bottomrule
      \end{tabular}
\caption{The hyper-parameters of CoMBD.}
\label{tab:appendix:CoMBD-params}
\end{table}

\subsection{Sub-Band Discriminator} \label{SBD}
The sub-modules of the SBD evaluate the sliced or the entire sub-band signals obtained by the PQMF analysis. Each sub-module comprises MDC layers in \citep{NeuralPhoto}, as shown in \cref{fig:avocodo_system}c. The MDC layer includes a convolution bank, a post-convolution layer, and a leaky ReLU activation layer. The convolution layers in the bank share kernel size but not dilation rates. \cref{tab:appendix:sbd-params} describes the hyper-parameters of SBD. The three sub-modules, i.e., tSBD, observe the changes in spectral features based on the time axis. We require each sub-module to cover a different range of frequency bands using different receptive fields. Consequently, the sub-module for low-frequency bands has large receptive fields, whereas the sub-module for broad frequency bands, it has small receptive fields.
\begin{table}[t]
\centering
      \begin{tabular}{lcccc}
        \toprule
             & tSBD \textsubscript{1}& tSBD\textsubscript{2} & tSBD\textsubscript{3} & fSBD \\
        \midrule
        Kernel size & 7 & 5 & 3 & 5  \\
        Kernel filters & \multicolumn{3}{c}{[64, 128, 256, 256, 256]} & \multicolumn{1}{c}{[32, 64, 128, 128, 128]}   \\
        Kernel stride & [1, 1, 3, 3, 1] &  [1, 1, 3, 3, 1] & [1, 1, 3, 3, 1] & [1, 1, 3, 3, 1]   \\
        Dilation factor  & [5, 7, 11]$\times$5 & [3, 5, 7]$\times$5 & [1, 2, 3]$\times$5 &  [1, 2, 3]$\times$3, [2, 3, 5]$\times$2 \\
        Sub-band range & [1:6] & [1:11] & [1:16] & [1:64] \\
        \bottomrule
      \end{tabular}
\caption{The hyper-parameters of SBD. The notation of sub-band range [$x$:$y$] means that $x$-th to $y$-th sub-band signals are selected.}
\label{tab:appendix:sbd-params}
\end{table}
\section{Pseudo Quadrature Mirror Filter bank} \label{PQMF}
Various types of digital filter banks are commonly used to inspect the frequency information of speech or image signals. Perfect decomposition and reconstruction are the key features of a digital filter. The quadrature mirror filter (QMF) bank is one of the most prevalent filters that can perfectly decompose and reconstruct a target signal. \citep{PQMF} proposed a pseudo-QMF (PQMF) bank based on the cosine modulated filter. The PQMF comprises an analysis and synthesis filter for multi-band sub-coding. In particular, PQMF has stopband attenuation over $100$dB, appropriate for sub-band coding without aliasing. Since it is possible to decimate speech signals without distortion, several neural vocoders employ the PQMF to achieve computational efficiency \citep{MB-MelGAN, DurIAN, Subband_Wavenet}. \cref{tab:appendix:PQMF-params} describes the hyper-parameters of PQMF based on the number of sub-bands. These values were selected empirically to prevent artifacts during decimation.
\begin{table}[h]
\centering
      \begin{tabular}{lcccc}
        \toprule
            & \multicolumn{4}{c}{Number of sub-bands} \\ \cmidrule{2-5}
             & 2 & 4 & 16 & 64 \\
        \midrule
        Taps & 256 & 192 & 256 & 256 \\
        Cut-off ratio & 0.25 & 0.13 & 0.03 & 0.1 \\
        Beta & 10.0 & 10.0 & 10.0 & 9.0 \\
        \bottomrule
      \end{tabular}
\caption{The hyper-parameters of PQMF.}
\label{tab:appendix:PQMF-params}
\end{table}

\appendix

\end{document}